\documentclass[10pt,letterpaper,twocolumn]{article} 
\usepackage{epstopdf}
\usepackage{ol2}
\usepackage[draft,implicit=false]{hyperref}
\usepackage{amsmath}
\usepackage{graphicx}

\begin{document}
\twocolumn[ 

\title{Spatial Mode Selective Waveguide with Hyperbolic Cladding}

\author{Y. Tang,$^{1,*}$ Z. Xi,$^{1,3}$ M. Xu,$^{1,2}$ S. B\"aumer,$^2$, A. J. L. Adam,$^1$ H. P. Urbach,${^1,^2}$}

\address{
$^1$ Optics Research Group, Department of Imaging Physics, Faculty of Applied Sciences, Delft University of Technology, 
\\Van der Waalsweg 8, 2628CH Delft, The Netherland\\
$^2$ The Netherlands Organisation for Applied Scientific Research~(TNO), \\
Stieltjesweg 1, 2628 CK Delft, Netherlands\\
$^3$ Z.Xi@tudelft.nl\\
$^*$ Corresponding author: Y.Tang-1@tudelft.nl
}

\begin{abstract}
Hyperbolic Meta-Materials~(HMMs) are anisotropic materials with permittivity tensor that has both positive and negative eigenvalues. Here we report that by using a type II HMM as cladding material, a waveguide which only supports higher order modes can be achieved, while the lower order modes become leaky and are absorbed in the HMM cladding. This counter-intuitive property can lead to novel application in optical communication and photonic integrated circuit. The loss in our HMM-Insulator-HMM~(HIH) waveguide is smaller than that of similar guided mode in a Metal-Insulator-Metal~(MIM) waveguide.
\end{abstract}
]

\noindent 
Meta-materials are structures engineered at the subwavelength scale to exhibit specific electromagnetic properties. The development of nanofabrication techniques allows to make these structures and to realize new properties that are unobtainable with conventional media. Among the varieties of meta-materials, Hyperbolic Meta-Materials~(HMMs) have gained tremendous attention. Their exotic hyperbolic dispersion property is the key to numerous emerging nano-photonics applications, including sub-diffraction-limit imaging~\cite{jacob2006optical,liu2007far,alu2014planarhyperlens,sun2015experimental,2016darkhyperlens,reviewer2_1,reviewer2_2}, Purcell factor enhancement~\cite{galfsky2015active,poddubny2013purcell,Belov2015purcell,sreekanth2014purcel, zhaowei2014enhancing, Capolino2014radiative}, sensing~\cite{vasilantonakis2015refractive, mackay2015sensing, tang2016tubular}, and waveguide engineering~\cite{he2012nanoscale,ishii2014plasmonic,babicheva2015finite,babicheva2015long,Zayats2015bulkwaveguide,reviewer1_1,reviewer3_1,reviewer3_2}. Here we report a new application of HMM for waveguide spatial mode engineering, which brings up new possibility in Spatial-Division Multiplexing~(SDM). 

SDM utilizes the last unexplored physical dimension, space, to further increase the data carrying capacity in optical communication~\cite{richardson2013space}. The excitation and separation of spatial modes is essential in SDM~\cite{chen2014compact}. In this letter, we propose a mode selective slab waveguide design by using a HMM as a cladding material. With a cladding consisting of HMM, the lower order modes with larger propagation constant become propagating wave in the HMM cladding material, such that they are turned into leaky modes. At the same time, higher order modes with smaller propagation constant are evanescent wave in the HMM cladding and remain guided in the core. Also by choosing the right parameter for the cladding one can design a 'single mode' waveguide only guiding one specific higher order mode, which can be applied as mode launcher or mode receiver in a SDM system. Compared to conventional spatial multiplexing techniques based on interference~\cite{riesen2013ultra} or holography~\cite{aoki2013holographic}, our approach merely modifies the waveguide property and requires no extra optical component, which is more compact and efficient.

\begin{figure}[htbp]
\centering
\fbox{\includegraphics[width=\linewidth]{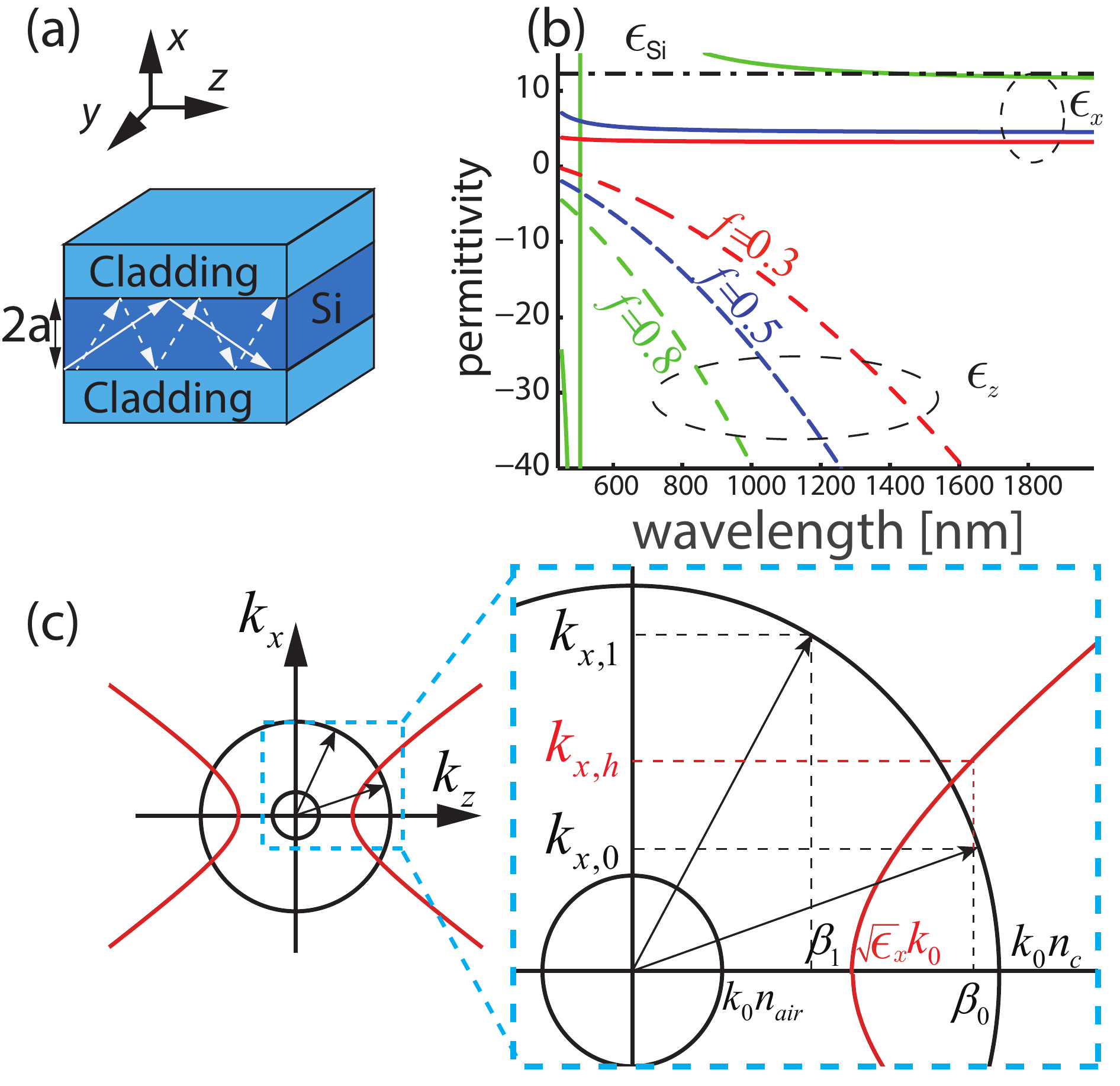}}
\caption{(a)~Schematic diagram of the multimode slab waveguide structure; (b)~Real part of the effective permittivity versus wavelength with filling factor $f$=0.3, 0.5 and 0.8. The horizontal black dash-dotted line indicates the permittivity of the silicon core; (c)~2D isofrequency contour of the isotropic core and the air~(in black) and the HMM cladding~(in red) in the $k_x$-$k_z$ plane; }
\label{fig:3d}
\end{figure}

We only consider transverse magnetic~(TM) wave~($E_y=0$) because the hyperbolic cladding only has the desired property for this state of polarization. The TM wave is propagating in a slab waveguide core towards the positive $z$ direction~(Fig.~\ref{fig:3d}a). The core thickness is $2a$. The magnetic field of the mode is independent of the $y$-coordinate, and has the form $H_y=H(x)\exp(i\beta_m z)$, where $\beta_m=k_{z,m}$ is the propagation constant of the $m^{\mathrm{th}}$ order mode along the $z$-direction, which satisfies the equation $\beta_m^2+k_{x,m}^2=n_ck_0^2$, where $k_{x,m}$ is the transverse wavenumber of the $m^{\mathrm{th}}$ order mode in the $x$ direction, $n_c$  is the refractive index of the core, and $k_0$ is the wavenumber in vacuum. The mode numbering convention in this paper is corresponding to the number of zeros of the magnetic field amplitude in transverse direction. The fundamental mode $TM_0$ has the smallest transverse wavenumber and thus corresponds to the largest propagation constant $\beta_0$. Higher order modes like $TM_1$, $TM_2$ with larger transverse wavenumber correspond to smaller propagation constant $\beta_1$, $\beta_2$, with $\beta_0>\beta_1>\beta_2$. For a conventional slab waveguide with sufficiently large index contrast or core thickness, such that the $m^{th}$ higher order mode is guided, all the lower order modes $TM_0$, ..., $TM_{m-1}$ are also guided. When the core thickness or the index contrast is reduced, the number of guided modes decreases in sequence of $m$. Therefore by decreasing the core thickness or index contrast one can select the lower order modes. However the selection of higher order modes can not be easily achieved through conventional method.

Now assume that the cladding material is replaced by a HMM, with its optical axis parallel to the $x$-axis. With respect to our Cartesian coordinate system, the permittivity tensor of the HMM is diagonal and given by:

\begin{equation}
\bar{\epsilon}= \left[ 
\begin{array}{ccc}
\epsilon_{x} &0 & 0 \\
0& \epsilon_{y} & 0 \\
0& 0 & \epsilon_{z} 
\end{array} \right].
\label{eq:tensor}
\end{equation}

If $\mathrm{Re}(\epsilon_y)=\mathrm{Re}(\epsilon_z)<0$ and $\mathrm{Re}(\epsilon_x)>0$, the HMM is of type II~\cite{caldwell2014sub}. For transverse electric~(TE) wave, the electric field only experiences the $\epsilon_y$ component and hence the HMM cladding behaves as a normal metallic material and the waveguide has the properties of a conventional waveguide described above. However, for TM waves, the projection of isofrequency contour on $k_x$-$k_z$ plane is a hyperbola, as indicated by the red curve in Fig.~\ref{fig:3d}c~(note $k_z$ is the tangential component of wave vector along the core/cladding interface). The intersection point of the hyperbola and the $k_z$-axis is given by $k_0\epsilon_x^{1/2}$. This number is the minimum value of $k_z$ for which waves are propagating in the HMM. For $k_z$ below this value, the wave is evanescent in the $x$-direction inside the cladding and hence for a guided mode there must hold: $\beta=k_z<k_0\epsilon_x^{1/2}$. If a lower order mode has a large propagation constant~(e.g. $\beta_0$ in Fig.~\ref{fig:3d}c), a propagating wave solution can be found in the HMM cladding, with transverse wavenumber $k_{x,h}$. Therefore the field becomes leaky. However, for a higher order mode with a smaller propagation constant~(e.g. $\beta_1<k_0\epsilon_x^{1/2}$ in Fig.~\ref{fig:3d}c), no solution for propagating wave can be found in the HMM cladding and therefore it is a guided mode. Hence a waveguide with HMM cladding can act as a mode selector that only allows higher order modes to propagate within the core.

Using the guided mode dispersion equation~\cite{ishii2014plasmonic}, the waveguide modes in HIH waveguide can be found. For a symmetric structure, the result solution is described by two sets of equation:
\begin{eqnarray}
&\mathrm{odd~modes}&: \tan(k_{x}a)=-\frac{k_{x}\epsilon_z}{\gamma_h\epsilon_c}, \\
&\mathrm{even~modes}&: \cot(k_{x}a)=\frac{k_{x}\epsilon_z}{\gamma_h\epsilon_c},
\label{eq:solution}
\end{eqnarray}
where $k_x=(n_c^2k_0^2-\beta^2)^{1/2}$ is the transverse wavenumber in the core, and $\gamma_h$ is the attenuation coefficient in the HMM cladding defined by $\gamma_h=[(\epsilon_z/\epsilon_x)\beta^2-\epsilon_zk_0^2]^{1/2}$. These equations are not only valid for oscillation mode~(photonic mode), but also valid for plasmonic mode with $k_x$ of imaginary value.

In our study, the waveguide core is made of silicon with refractive index 3.5 for all wavelengths we considered in this paper. Though no natural hyperbolic material around 1550~nm wavelength is found so far~\cite{nature2015compendium}, the HMM cladding can be approximated by a multilayer structure consisting of periodically alternating metal and dielectric layers, of which the interfaces are parallel to the $y-z$ plane. According to the effective medium theory~(EMT)~\cite{cai2010optical}, when the thickness of each layer is much smaller than the wavelength, the eigenvalues of the effective optical permittivity tensor are given by $\epsilon_{x}^{-1}=f\epsilon_m^{-1}+(1-f)\epsilon_d^{-1}$, $\epsilon_{y}=\epsilon_{z}=f\epsilon_m+(1-f)\epsilon_d$, where $f$ is the filling factor of metal in a unit cell, $\epsilon_{m}$ and $\epsilon_{d}$ are the permittivity of the metal and the dielectric, respectively. We use $\mathrm{Ag}$ and $\mathrm{Al_2O_3}$ as the constituent materials, with refractive index of $\mathrm{Al_2O_3}$ equals to 1.5 for all wavelengths, and the permittivity of $\mathrm{Ag}$ taken from~\cite{johnson1972optical}. The thickness of the core is chosen as $2a$ = 1000~nm. Because the metal is lossy, the propagation constant $\beta$ is complex. We define the effective mode index $n_{\mathrm{eff}}$ and the propagation length $L$ as:
\begin{align}
n_{\mathrm{eff}}&=\mathrm{Re}(\beta)k_0^{-1},\\
L&=[\mathrm{Im}(2\beta)]^{-1}.
\label{eq:pg}
\end{align}

In Fig.~\ref{fig:3d}b, the real parts of the effective $\epsilon_x$ and $\epsilon_z$ are shown as a function of wavelength. In a broad wavelength range, $\epsilon_{x}$ is positive while $\epsilon_{z}$ is negative, giving a type II hyperbolic behavior. By changing the filing factor $f$, the wavelength range of the hyperbolic dispersion and the permittivity values can be tuned. The value of $\epsilon_x$ increases as the filling factor increases. In order to fulfil the requirement ${\epsilon_x}^{1/2}<n_c$, the filling factor must have a moderate value. We choose a filling factor of 0.5, leading to $\mathrm{Re}(\epsilon_x^{1/2})\approx2.3<n_{\textrm{Si}}$.

\begin{figure}[htbp]
\centering
\fbox{\includegraphics[width=\linewidth]{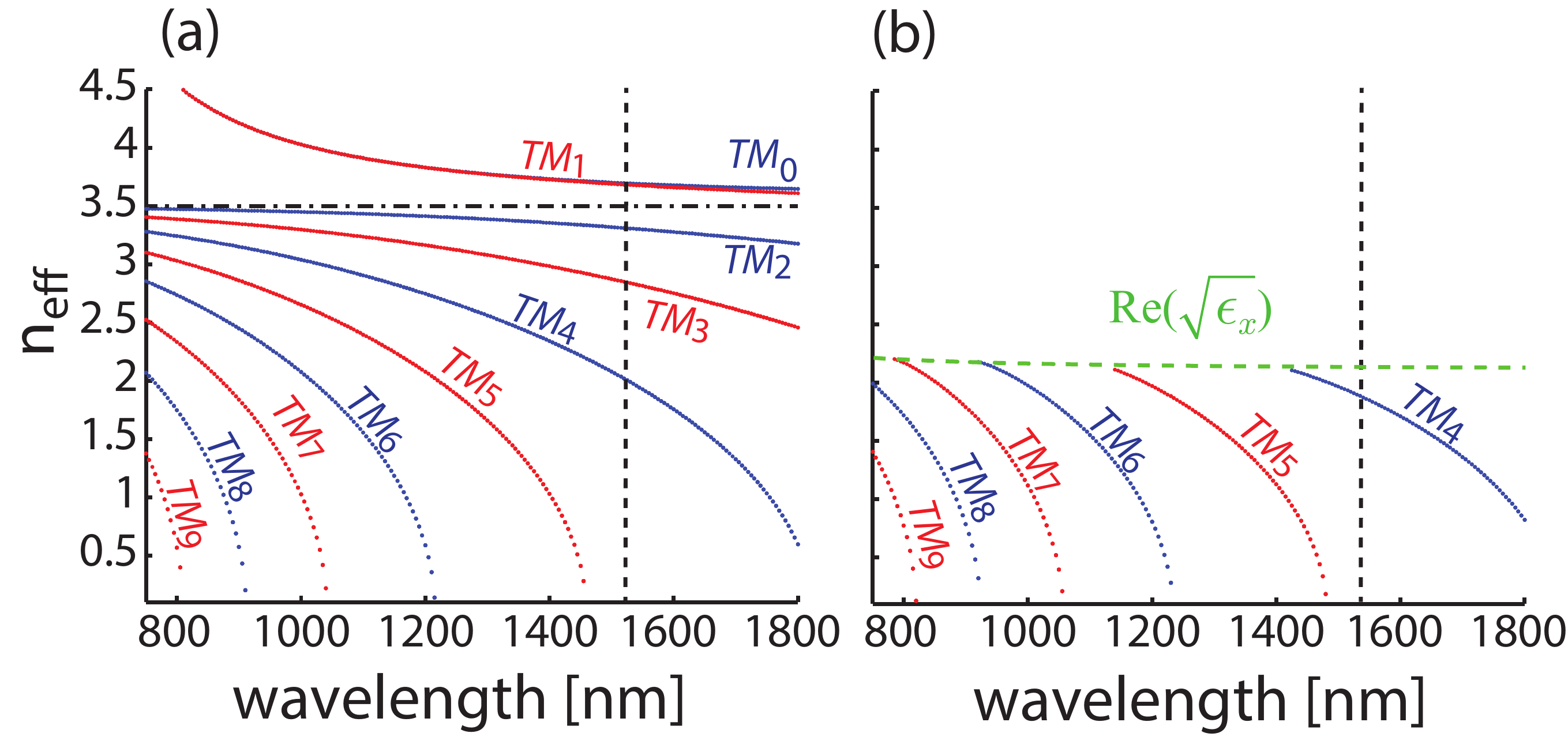}}
\caption{Effective indices of guided modes for different filing factors: (a) $f$=1~(MIM); (b)$f$=0.5. The waveguide thicknesses are both $2a=1000$~nm. The horizontal dash-dotted line indicates the refractive index of the core and the vertical black dashed line indicates the wavelength of interest~(1550~nm).}
\label{fig:mode_f}
\end{figure}

For these values of the parameters, the solutions of the dispersion relations (2) and (3) of the guided TM modes are shown in Fig.~\ref{fig:mode_f}b. For comparison, the MIM~(metal-insulator-metal) case is shown in Fig~\ref{fig:mode_f}a, obtained by setting the filling factor equal to 1. The red curves and the blue curves represent the odd and even modes, respectively. As can be seen in Fig.~\ref{fig:mode_f}a, when $f=1$~(MIM), all the guided modes are found. At the wavelength 800~nm, 10 guided modes are supported by the waveguide. The $TM_0$ and $TM_1$ are the symmetric and anti-symmetric plasmonic mode, respectively, and the other higher order modes are oscillatory modes~\cite{name2012multimode}. The $TM_0$ mode always exists for all wavelength. But for other orders of modes, as the wavelength increases~(frequency decreases), higher order modes reach their cut-off frequency first and disappear. Eventually for $\lambda$=1600~nm, only the first five order modes remain~($TM_0$ to $TM_4$). However, in the case of HIH waveguide ($f=0.5$) in Fig.~\ref{fig:mode_f}b, the first four order modes from $TM_0$ to $TM_3$ do not exist at the wavelengths shown in the plot. The lowest guided mode varies from $TM_4$ to $TM_9$ when the wavelength is decreased. The modes with $n_{\mathrm{neff}}>Re({\epsilon_x}^{1/2})$ above the green dashed line disappear, because they propagate in the HMM cladding. If the filling factor is further decreased, the lowest order of the guided modes increases. 

\begin{figure}[htbp]
\centering
\fbox{\includegraphics[width=\linewidth]{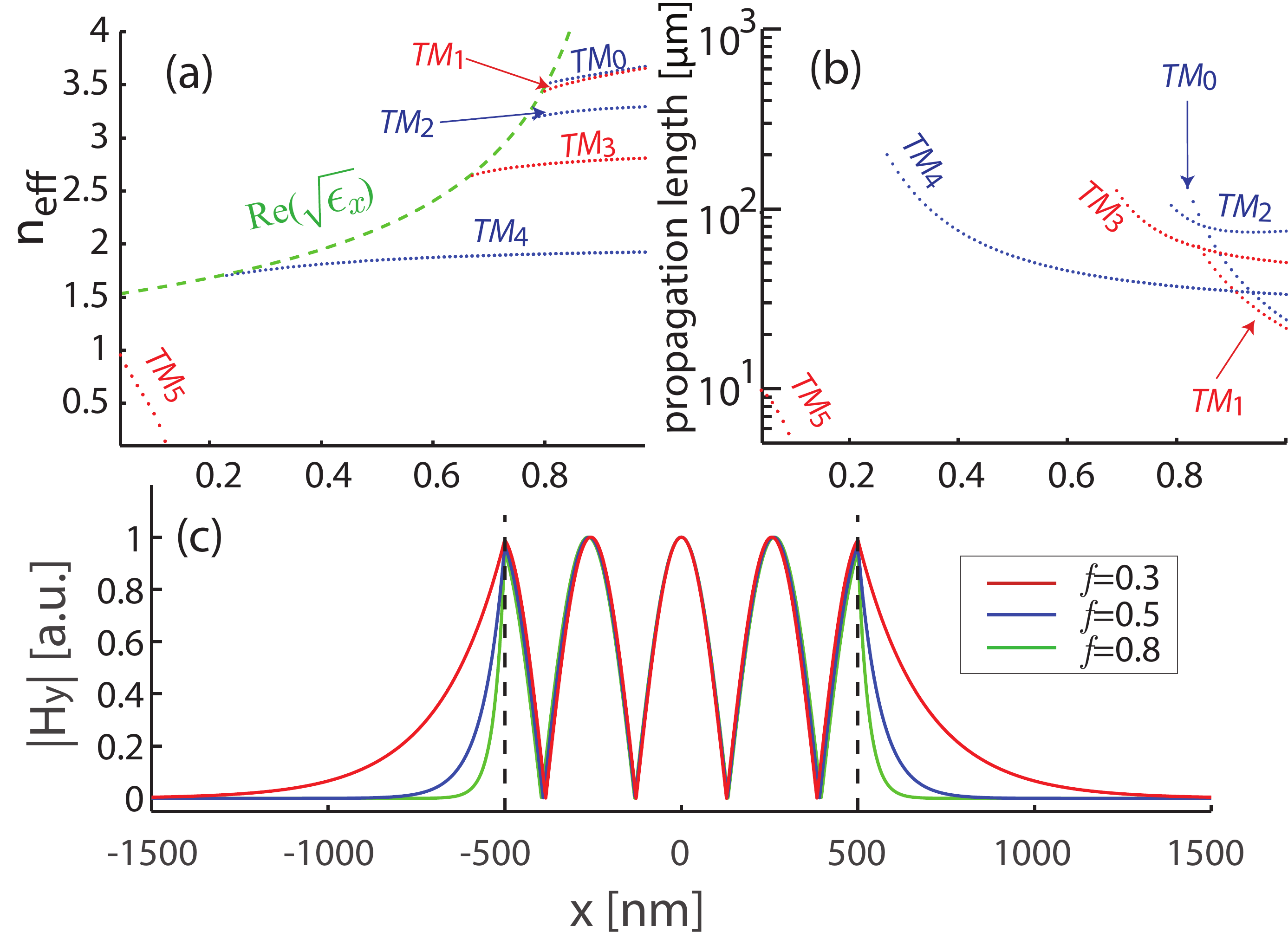}}
\caption{Effective indices (a) and propagation length (b) of guided modes as a function of the filling factor $f$ at $\lambda$=1550~nm and waveguide thickness $2a=1000$~nm. (c)~Magnetic field amplitude distribution of the $TM_4$ mode with different filling factors. Vertical dashed line indicates the waveguide boundary.}
\label{fig:1550mode_f}
\end{figure}

The mode selection property is further studied as a function of the filling factor for the hyperbolic cladding at 1550~nm wavelength. Both the effective mode index and the propagation length are shown in Fig.~\ref{fig:1550mode_f}. For all effective indices shown in Fig.~\ref{fig:1550mode_f}a, five modes exist for $f$=1~(MIM). As the filling factor decreases, the plasmonic mode $TM_0$ and $TM_1$ are cut-off first at around $f=0.8$. The $TM_2$ mode only exists for $f>$0.77, the $TM_3$ exists for $f>$0.67, and the $TM_4$ mode disappears only when $f<$0.22. The $TM_5$ emerges for $f<$0.12. This is because the cut-off frequency of the $TM_5$ mode red-shifts when the filling factor decreases. Again, the effective index can not exceed $\mathrm{Re}({\epsilon_x}^{1/2})$, which is indicated by the green dashed line. It should be pointed out that for $f<0.67$, the waveguide becomes a single mode waveguide which only guides either the $TM_4$ mode or the $TM_5$ mode. Such phenomenon can not be achieved with conventional waveguides. It is seen in Fig.~\ref{fig:1550mode_f}b that the propagation length of all modes are longer for smaller filling factor. The reason is that with a smaller filling factor, less metal is used in the HMM cladding, and thus the modes experience less metallic loss. For the $TM_4$ mode, the maximum propagation length can exceed 400~µm, which is an order of magnitude longer than the propagation length of this mode in a metallic waveguide with Ag cladding~(i.e. with fill factor of 1). However, in the HIH case, the field confinement becomes relatively weak. As shown in Fig.~\ref{fig:1550mode_f}(c) the field has the longest decay tail in the transverse direction for $f$=0.3. This is because the field decay rate in the cladding is determined by $\gamma_h$, and when $\beta$ approaches $\epsilon_x^{1/2}k_0$, the value of $\gamma_h$ becomes smaller, and hence the field distribution is broadened. 

\begin{figure}[htbp]
\centering
\fbox{\includegraphics[width=\linewidth]{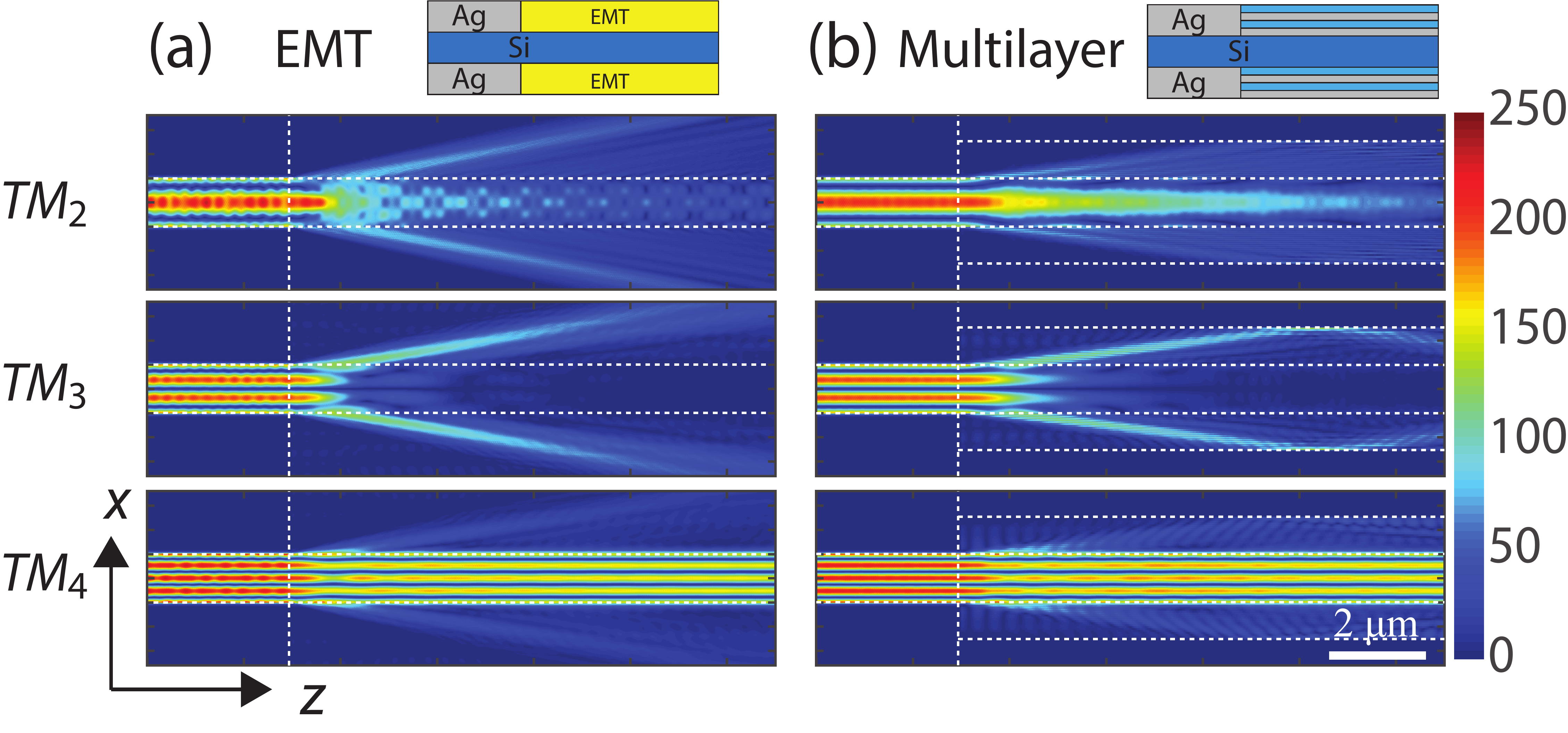}}
\caption{Simulated magnetic field amplitude of three guided modes in a MIM waveguide coupled to a HIH waveguide. The filling factor of the multilayer is 0.5. The white dashed line indicates the boundary. (a) Hyperbolic cladding modelled using the effective medium theory ; (b) Multilayer cladding. Inset: schematic of the waveguide.}\label{fig:simMulti}
\end{figure}

To confirm the validity of this particular mode selection property, both the HMM with effective medium model and the real multilayer HMM cladded waveguide are simulated for the wavelength of 1550~nm using the finite element method~(FEM) software COMSOL~4.4. Full losses is considered in the simulation. In the effective medium situation, the HMM cladding is modeled as a semi-infinite half-space. In the real multilayer case, the layers are made of 20 pairs of Ag and $\mathrm{Al_2O_3}$ on each side, of which Ag is next to the core. The thicknesses of the dielectric and the metal layers are both 20~nm, hence the filling factor is 0.5. The material outside the cladding is air~(n=1). In both simulations, a MIM waveguide with Ag as metal~(on the left) is connected to a HIH waveguide~(on the right) with identical silicon core~(2a = 1000~nm, $n_c = n_{\mathrm{Si}} =3.5$), as shown in Fig.~\ref{fig:simMulti}. The $TM_2$, $TM_3$ and $TM_4$ modes are excited separately in the MIM waveguide and enter from the left end. If the mode is also supported by the HIH waveguide on the right, the energy will be coupled at the interface and continues to be confined in the core of the HIH waveguide, otherwise it will leak out into the cladding layer. 

The results are shown in Fig.~\ref{fig:simMulti}. From the top to the bottom, the magnetic field amplitude of the $TM_2$, $TM_3$ and $TM_4$ modes are shown. In Fig.~\ref{fig:simMulti}a, where the effective medium is used, the field of the $TM_2$ and the $TM_3$ mode leak quickly into the cladding after entering the HIH region, and the field amplitude decreases fast. However for the $TM_4$ mode most of the field is still confined in the HIH waveguide, because it is a guided mode in HIH waveguide. A small leakage and reflection is observed due to a small mode mismatch between the MIM and the HIH waveguide. For the multilayer cladding shown in Fig.~\ref{fig:simMulti}b, the field of the $TM_2$ mode spreads out after entering the HIH region, but not as quickly as in the EMT case. The propagation length of $TM_2$ is only 3~µm. The field of the $TM_3$ mode immediately starts to leak into the cladding, and reflects back at the outer interface between the cladding and the air because of mode mismatch, and dissipates in the cladding. The field of the $TM_4$ mode is well confined to the core of the HIH waveguide and only a small portion of energy leaks into the cladding. The simulation result confirms that this HIH waveguide possesses high-pass mode selection property. For multilayer hyperbolic cladding, further study is still needed to determine the minimum number of needed layers in our application. However, in \cite{babicheva2015finite}, it is shown that, for plasmonic modes, when the number of periods exceeds ten, the exact calculation can closely match the effective medium results.

To summarize, we propose a novel mode selective waveguide by using a HMM as cladding material. By choosing the right dielectric and metal and the proper filling factor for HMM, we can achieve a waveguide structure which only allows certain higher order modes to be guided. Moreover, the propagation length of HIH waveguide can be an order of magnitude longer than that of a MIM waveguide. Our design combining meta-material with conventional waveguide design, opens up new possibilities for waveguide engineering.

\section*{Acknowledgments}
Y. Tang acknowledges the financial support from China Scholarship Council~(CSC) and Nederlandse Organisatie voor Toegepast Natuurwetenschappelijk Onderzoek~(TNO). 


\end{document}